\documentclass[twocolumn,secnumarabic,amssymb, nobibnotes, aps, prd]{revtex4-1}
\usepackage{graphicx}
\usepackage{dcolumn}
\usepackage{bm}
\usepackage{comment}
\usepackage{epstopdf}
\usepackage{natbib}

\newcommand{\be}{\begin{equation}}
\newcommand{\ee}{\end{equation}}
\newcommand{\bea}{\begin{eqnarray}}
\newcommand{\eea}{\end{eqnarray}}

\begin{document}

\preprint{AIP/123-QED}

\title{Evidence of a two-stage melting of Wigner solids in two dimensions}

\author{Jian Huang\thanks{email:jianhuang@wayne.edu}}
\affiliation{Department of Physics and Astronomy, Wayne State University, Detroit, MI 48201, USA}
\author{Talbot Knighton}
\affiliation{Department of Physics and Astronomy, Wayne State University, Detroit, MI 48201, USA}
\author{Zhe Wu}
\affiliation{Department of Physics and Astronomy, Wayne State University, Detroit, MI 48201, USA}
\author{Alessandro Serafin}
\affiliation{National High Field Magnetic Laboratory, Tallahassee, FL 32310}
\author{J. S. Xia}
\affiliation{National High Field Magnetic Laboratory, Tallahassee, FL 32310}
\author{L. N. Pfeiffer}
\affiliation{Department of Electrical Engineering, Princeton University, Princeton, NJ 08544}
\author{K. W. West}
\affiliation{Department of Electrical Engineering, Princeton University, Princeton, NJ 08544}

\date{\today}
             
\begin{abstract}
\end{abstract}

\pacs{Valid PACS appear here}
\maketitle
 
Two-dimensional (2D) solid-liquid transition (SLT)~\cite{Mermin1966Absence,Mermin1968Crystalline,Kosterlitz1972Long} concerns fundamental concepts of long-range correlations vital to magnetism, superconductivity, superfluidity, and topological matters. A long sought-after example is the melting of a Wigner Crystal (WC)~\cite{Wigner1934Interaction} of electrons. Detection efforts have targeted distinctive collective modes such as pinning by disorder, resonant-frequency absorption of Goldstone modes, and melting transition. However, only one-step second-order melting of softly-pinned modes was reported. Without the evidence of genuine pinning as exhibited in the charge density waves (CDWs)~\cite{PinningCDW}, these modes are likely intermediate phases which are only part of a complete SLT. To verify if there is a WC-intermediate phase transition will not only provide a solid proof of a WC, but will also unveil the nature of the SLT in relation to the two-stage Kosterlitz-Thouless (KT) model~\cite{Kosterlitz1972Long,Kosterlitz1973Ordering,Halperin1978Theory,Nelson1979Dislocation,hexatic_ceperley,hexatic_nelson, Young1979Melting}. 
Through transport studies of ultra-dilute high-purity 2D systems, this work presents evidence for not only a WC, but also for a two-stage WC-liquid SLT mediated by a first-order WC-intermediate phase transition.

A WC is expected in ultra-dilute systems where the ratio of the inter-particle Coulomb energy $E_{ee}$ and the Fermi energy $E_F$, $r_s=E_{ee}/E_F=a/a_B$, surpasses 37~\cite{Tanatar1989Ground}. $a=1/\sqrt{\pi n}$ is the Wigner-Seitz radius for electron density $n$ and $a_B=\hbar^2\epsilon/m^{*} e^2$ is the Bohr radius. The smallness of $E_{ee}=e^2/\epsilon a\sim1 me$V and $E_F$ 
$=n\pi\hbar^2/m^*\sim25\mu e$V makes a WC fragile against random disorders which tend to render an Anderson insulator~\cite{Anderson1958Absence} or a glass. More subtly, disorder fluctuations, along with the quantum fluctuations, 
reduce the melting temperature $T_m$ well below the classical melting point $T_{cm}\approx e^2\sqrt{\pi n}/\epsilon\Gamma$ ($\Gamma\approx130$). As $T_{cm}$ is $\sim100$~mK for $r_s=37$, $T_m$ can be even lower than the typical cooling capability of 10-30~mK depending on the disorder level. Disorder suppression is therefore key to keeping $T_m$ accessible. As conventional methods cannot provide cooling of the carriers below such a small $T_m$, we adopt a novel immersion cell cooling technique and demonstrate an actual $T_m\sim30$~mK even in ultra-pure systems. The widely reported $T_m$, $\sim$150 and 300mK, is actually the liquefaction point of a part of a complete transition.  

Experimental studies of WCs primarily employ transport and absorption techniques since a direct scanning method is not yet realistic. The most studied are the reentrant insulating phases (RIPs) in the fractional quantum Hall regime where AC and DC~\cite{Jiang1991Magnetotransport,Goldman1990Evidence,Williams1991Conduction}, resonant absorption (rf, microwaves, and acoustic waves)~\cite{RFAndrei,Chen2006Melting,Zhu2010Observation,acousticWC}, and tunneling~\cite{jang2016sharp} techniques are employed to detect the collective responses of putative WCs under pinning. Though a pinned WC should exhibit a sharp nonlinear threshold DC response, a benchmark for pinned CDWs~\cite{Gruener1988dynamics}, most are reported as soft and rounded nonlinear IVs with small pinning strength ($\leq 1M\Omega$) and a large variance in the obscurely-defined pinning thresholds ($\sim$100s of mV/cm). Activated $T$ dependent pinning~\cite{DIorio1992Reentrant} hints that these putatively pinned states, with an exponentially decreasing translational correlation length $\xi$, are actually intermediate/mixed phases~\cite{Nelson1980Solid}. This includes the systems used for resonant-absorption studies where collective modes are identified~\cite{RFAndrei,Chen2006Melting,Zhu2010Observation,acousticWC,jang2016sharp}. In contrast, $T$ dependence in a pinned WC should be moderate as quantum effects dominate. 


On the other hand, the more desired $B=0$ studies require ultra-dilute carrier concentrations, i.e. $\sim1\times10^{9}$ cm$^{-2}$ for electrons or $\sim4\times10^{9}$ cm$^{-2}$ for holes in {\it GaAs} systems, in order to reach $r_s=37$. These have not been achieved by previous studies~\cite{Kravchenko1991Two,Pudalov1993Zero,Yoon1999Wigner} due to a greater challenge associated with a rising disorder effect caused by weakened inter-particle screening~\cite{Huang2011Light}. This is why localization occurs even in fairly clean systems when $a\propto n^{-1/2}$ approaches the localization length. We have overcome this barrier by realizing ultra-dilute 2D hole systems with much suppressed disorder.

Key observations include rigorously-pinned WCs and a first-order two-stage SLT. Enormous pinning with hundreds of M$\Omega$ resistivity is observed at $T<T_m\approx30$ mK via sharp threshold DC-IV marked by more than three orders of magnitude drop in the differential resistance $r_d$ across a threshold point. Pinning collapses sharply beyond $T_m$, $\sim (1/4)T_{cm}$, until a liquid phase is reached. The presentation is divided into two parts: The first is a study of the RIP near filling $\nu = 1/3$ using $p$-doped quantum wells, and the second is a zero-$B$-field study of ultra-dilute holes in undoped heterojunction-insulated-gate field-effect transistors (HIGFETs) at $r_s\geq$ 40. 

\begin{figure}[h]
\vspace{-0pt}
\includegraphics[totalheight=1.35in]{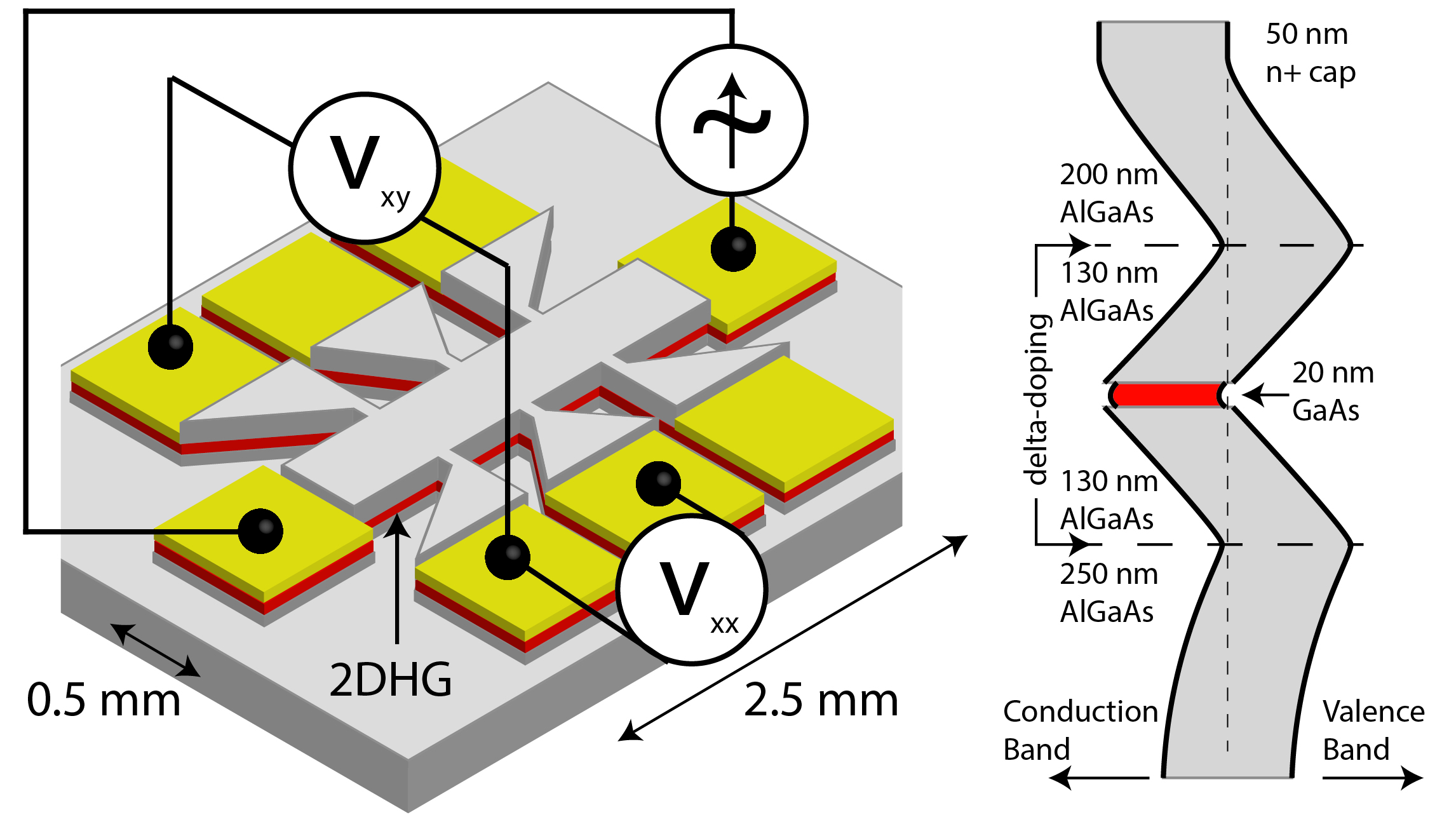}
\caption{\label{fig:sample} (a) Sample dimensions and Hall measurement configuration. (b) Band diagram of the quantum square well.}
\vspace{-10pt}
\end{figure}
The samples used for the RIP measurement are lightly doped $p$-type (100) GaAs quantum square wells. The density $p$ is $\sim4\times10^{10}$~cm$^{-2}$ ($r_s=24$), with mobility of $\mu \approx 2.5\times10^{6}$ cm$^2$/Vs. $a=1/\sqrt{\pi p}\sim 28 n$m is half of the average carrier spacing, and $m^{*}\approx0.45 m_e$ is effective mass. The sample geometry is a 2.5$\times$0.5mm Hallbar realized via a wet chemical etch. Thermally deposited AuBe 
pads annealed at $460~^{\circ}$C achieve excellent Ohmic contacts to the 2D carriers, with measured contact resistances $\sim400~\Omega$. 
Measurements are performed in a dilution refrigerator inside a fully shielded room, allowing the electronics to perform at their specified ratings.

\begin{figure}[b]
\includegraphics[totalheight=1.5in]{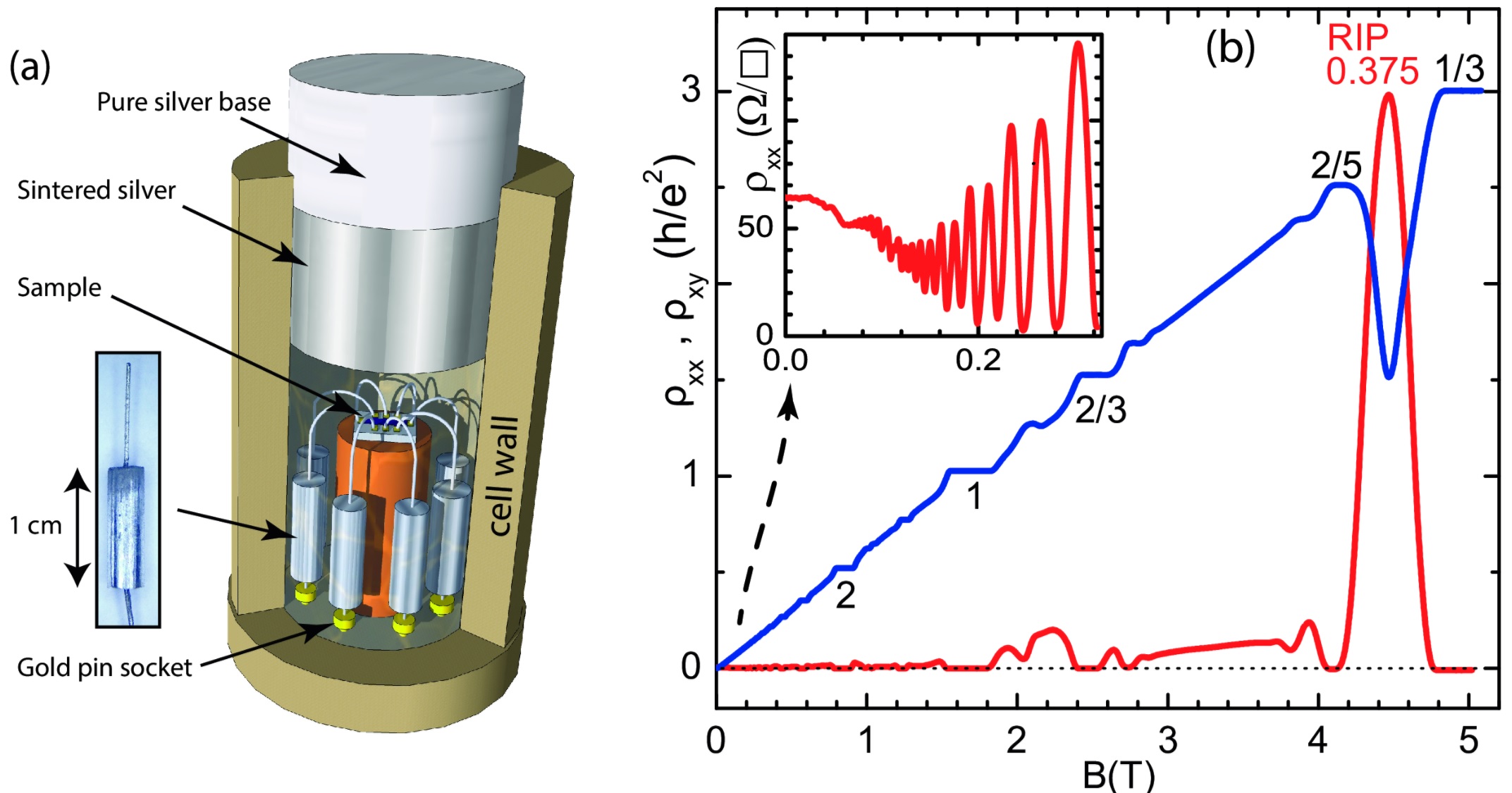}
\vspace{-5pt}
\caption{\label{fig:RIP} (a) Cooling schematics inside a helium-3 immersion cell. (b) Magnetoresistance and Hall resistance obtained at 10~mK. Inset: SdH oscillations.}
\vspace{-10pt}
\end{figure}

Cooling carriers to $T_{bath}$ is critical and challenging, and it must be achieved via sufficient heat exchange which conventional methods can not accomplish. We adopt well-designed helium-3 sample immersion cells [Fig.~\ref{fig:RIP}(a)], which is the proven method for effective cooling down to 5 mK~\cite{Huang2007Disappearance}, and achieve thermalisation of the 2D holes with the mixing chamber (mc). The polycarbon cell is vacuum-tight and mounted at the lower end of a silver cold finger with its top fastened to the mc plate. The roof is a sintered silver cylindrical-block extension of the cold finger made by compressed pure silver micro-particles. During operation, helium-3 gas is continuously fed through a capillary into the cell where it condenses to fill the volume completely. Saturated sintered silver block provides $\sim$30 m$^2$ contact area to cool the helium-3 bath. Major cooling of the 2D holes is realized via efficiently heat-sinking the metal contacts through sintered silver pillars providing 2.5~m$^2$ surface area per lead. $T$ is monitored through a helium-3 melting curve thermometer inside an identical cell located next to the sample cell. The $T$ differential between the bath and the mc is $\leq$0.1mK at all times.


\begin{figure}[b]
\vspace{-10pt}
\centering
\includegraphics[width=3.3in]{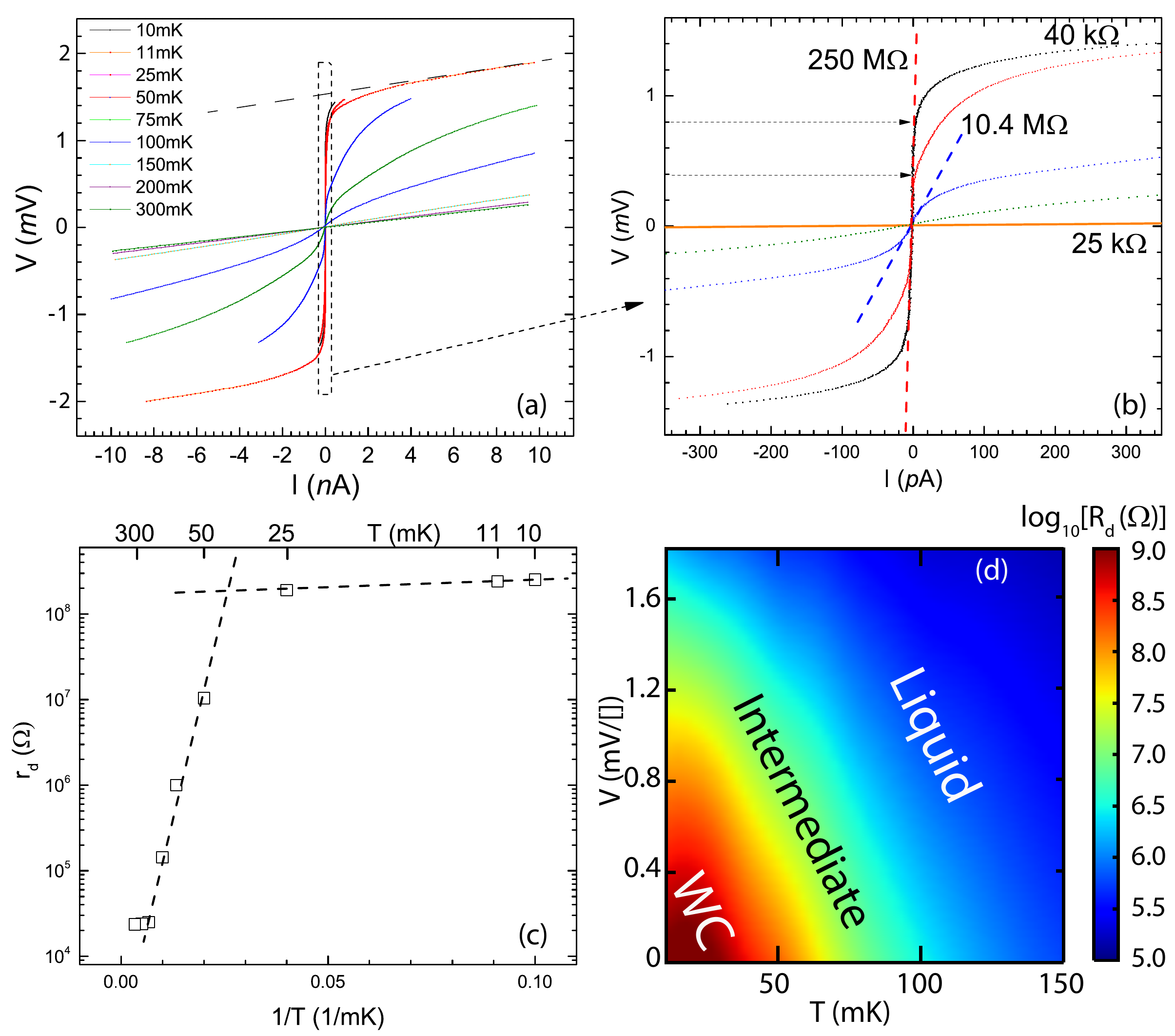}
\vspace{-10pt}
\caption{\label{fig:pinning} Color online (a) DC-IV curves for various $T$ measured at $B=$4.5~T. (b) Amplified view of the dotted box in (a). (c) Piecewise $T$ dependence of maximum $r_d(T)$ on semi-logarithmic scales. Dotted lines are guide to the eye. 
(c) Colored contour phase diagram based on $\log_{10}{R_d}(T)$ values.}
\end{figure}

Fig.~\ref{fig:RIP} (b) shows the magnetoresistance MR ($\rho_{xx}$) and the Hall resistance ($\rho_{xy}$) obtained at 10 mK via a four-terminal AC technique. 
The inset shows the Shubnikov de Haas (SdH) oscillations starting at 0.05~T. The RIP peak centers at $B = 4.5~$T ($\nu =0.375$) between fillings $\nu = 2/5$ and 1/3, with a dip in $\rho_{xy}$ consistent with previous studies~\cite{Sajoto1993Hall}. Fixing $B=$4.5~Tesla, where the magnetic length $l_B=\sqrt{hc/eB}\approx a\sim28$nm, an electrometer-level DC-IV setup is adopted to obtain the IV relationship for several $T$ from 10 to 300mK [Fig.~\ref{fig:pinning}(a)], with a voltage bias $V$ between $\pm$10mV (at $0.1\mu$V resolutions) to cover sufficient dynamic range. Current sensing is realized via a low-noise preamp at 50 $f$A precision. The $T$ dependence reveals three types of IV characteristics as a function of $T$: sharp threshold below some critical $T\sim$40 mK; rounded/soft nonlinear IV between 40 and 120 mK; and linear IV beyond 120~mK. 

For a fixed $T$ below 40~mK, the dynamical response shows a sharp threshold IV with a striking six-thousand-times plunge in the differential resistance $r_d=dV/dI$, from $\sim$250 M$\Omega/\square$ down to $\sim$40k$\Omega/\square$. Joule heating of $<10^{-15}$ Watts at $V_c$ is negligible. The enormous sub-threshold $r_d$ presents a greater pinning strength than reported for pinned CDWs~\cite{Gruener1988dynamics}. Sharply rising $V$ in the absence of current flow (with a current less than the threshold current $I_{c}\sim2-3p$A) is consistent with a pinned crystal whose potential energy grows via elastic deformation on a scale of $\xi$ in response to an increasing external field $E$. Conventionally, the threshold is thought to correspond to a pinning and depinning of a WC~\cite{WC_helium}. Indeed, the supra-threshold $r_d(V)$ becomes linear above $V=1.5$~mV, consistent with a non-pinned state. However, the values for the linear supra-threshold $r_d(V)$, $\sim$40k$\Omega/\square$, are in excellent agreement with that of a liquid state above the liquefaction point (shown later), indicating a pressure ($E$)-driven melting. It implies an intermediate phase between the pinned WC and a liquid. As this region is where $E_c$ was usually assigned in previous studies, more detailed examination is necessary. 

Fig.~\ref{fig:pinning}(b) is a zoom-in view of the (dotted line) box of Fig.~\ref{fig:pinning}(a) in which IVs are shown for 10, 25, 50, 75 and 150~mK. For the two IVs for 10 and 25~mK, $r_d$ for pinned modes, $\sim250$M$\Omega$, is mostly identical up to $V=0.4$~mV. It confirms only a slight $T$ dependence of $\xi$ below 40~mK. This lack of $T$ dependence is in sharp contrast to previous results~\cite{Jiang1991Magnetotransport,Goldman1990Evidence,Williams1991Conduction,RFAndrei,Yoon1999Wigner,Chen2006Melting,Zhu2010Observation,acousticWC,jang2016sharp,Kravchenko1991Two,Pudalov1993Zero,Yoon1999Wigner} and
is consistent with a quantum crystal behavior mentioned earlier. Now, deviation from pinning occurs at different points for different $T$, 0.4~mV (for 25 mK) and 0.8~mV (for 10 mK), beyond which are the intermediate phases characterized by soft/rounder IVs. Now, $E_c=V_c/L$ ($L$-length over $V_c$ is applied) is naturally defined as the deviation point. Lower $E_c$ for $T=25$~mK is qualitatively consistent with the Lindemann criterion~\cite{hexatic_nelson} for higher $T$. 

However, for $T$ above 40 mK, $E_c$ approaches zero and rigorous pinning disappears in the intermediate phase. Rounded/soft nonlinear IVs dominate even at zero-bias. Nevertheless, substantial $r_d$, i.e. $\sim10$M$\Omega$ at 50~mK, indicates a finite shear modulus $\kappa$, sustained by collective modes, only expected for a first-order transition\cite{Littlewood1994}. The disappearance of $E_c$ is consistent with a discontinuous $\xi$ across $T\sim$40~mK. It is now clear that the rounded nonlinear IV, which has been regarded by several previous studies ~\cite{Goldman1990Evidence,Kravchenko1991Two,Pudalov1993Zero,Yoon1999Wigner} as pinned WCs, is actually of an intermediate phase, which crosses over to a liquid in a second-order manner at higher $T$. The linear IV at 150 mK is beyond the liquefaction point ($T_l$). 

$T$-driven melting transition is captured by the maximum $r_d(T)$ found at minimal external bias for different $T$. Fig.~\ref{fig:pinning}(c) shows a pronounced piece-wise dependence across a critical temperature of $\sim38$~mK defined as $T_m$. A discontinuous $\xi$ acoss $T_m$, thus also carrier internal energy, is now apparent. Determination of $\xi$ is difficult~\cite{Lee1978Dynamics,Fukuyama1978Dynamics} and we provide only a rough estimate which turns out to be substantially larger than previous reports\cite{Goldman1988Evidence,Williams1991Conduction,jang2016sharp}. The maximum potential energy at $E_c$ is $U=Nw$. $w=eE_c a\sim 0.024\mu$V$\ll T$ is the single particle potential energy. $N=p\xi^2$ is the number of carriers on a scale of $\xi$. For $T=25m$K, $E_c=V_c/L\sim8$ $mV/$cm where $L=0.5$~mm. By balancing the electrical force $NeE_c$ with the pinning force $\kappa a$ as shown in reference~\cite{Goldman1990Evidence,Williams1991Conduction} ($\kappa$ being the shear modulus), $N\sim 1.5\times10^5$ or $\xi\geq 10\mu$m is obtained. $U$ is $\sim2.4me$V (or 30~K), comparable to $E_{ee}$. A complete phase diagram on $T$-$V$ axes is shown in Fig.~\ref{fig:pinning}(c). 

To identify the nature of the intermediate phase is difficult because the relationship between $r_d$ and $\xi(T)$ is yet to be proven. Here, we show, as a minor point, that $r_d(T)$ can be fitted to $r_d=r_0\exp[c/(T-40mK)^\gamma]$ with $r_0\approx23$ and $c\approx 9.5$, in the same trend as the exponentially decreasing $\xi(T)$ modeled for a hexatic phase~\cite{Nelson1979Dislocation}: $\xi\sim\exp[c/(T-T_m)^\gamma]$ ($\gamma\approx0.3696$). 

\begin{figure}[t]
\vspace{-0pt}
\includegraphics[totalheight=1.3in]{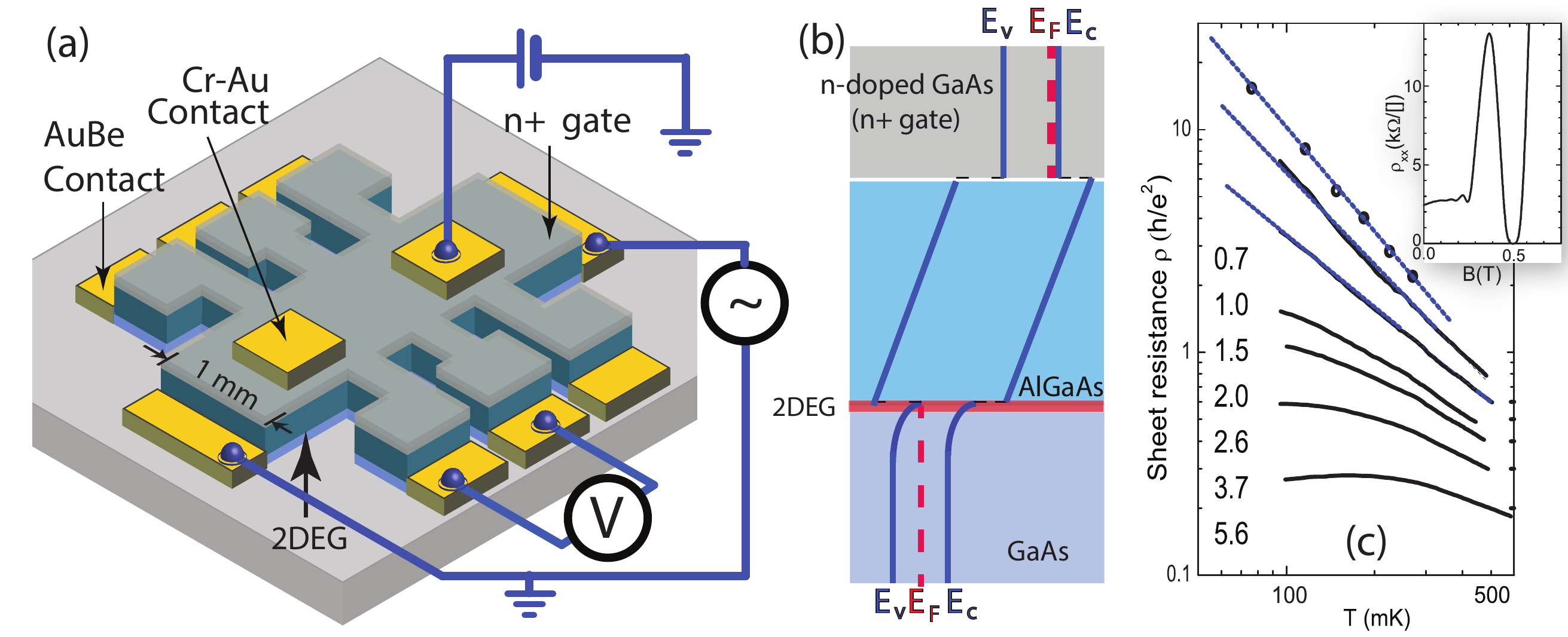}
\vspace{-5pt}
\caption{\label{fig:HIGFET} (a) HIGFET sample and measurement schematics. (b) Band diagram showing accumulation of holes. (c) Power law $T$ dependence of $\rho(T)$ on $\log-\log$ scales for indicated $p$'s ($\times10^{9}$ cm$^{-2}$). Inset: $\rho_{xx}(B)$ for $p=1.2\times10^{10}$ cm$^{-2}$.}
\vspace{-10pt}
\end{figure}

We now turn to the zero-field study for which undoped GaAs/AlGaAs HIGFETs~\cite{Kane1995High,Noh2003Interaction,HUANG2007TWO} are adopted for the benefit of more suppressed disorder than any doped devices. A 6\,mm$\times 0.8$\,mm Hall bar [Fig.~\ref{fig:HIGFET}(a)] is realized with a self-align fabrication process~\cite{HUANG2007TWO}. 
The carrier accumulation at the hetero-interface is capacitively induced  through biasing a top gate beyond a turn-on voltage, $\sim -1.3$~V, at which the valance band edge meets the chemical potential [Fig.~\ref{fig:HIGFET}(b)]. The band gap of the 600nm-thick $Al_{0.3}Ga_{0.7}As$ barrier is $\sim$2 $e$V. Owning to the superior crystal quality, zero gate leakage (not measurable on a 0.05 $p$A scale) is achieved at all operating bias. $p$ is tunable from $4\times10^{10}$ down to $7\times10^{8}$ cm$^{-2}$. Inset of Fig.~\ref{fig:HIGFET}(c) is a typical $R_{xx}$ for $p\sim1\times10^{10}$ cm$^{-2}$.

Accessing $r_s\geq 37$ requires $p\leq4.2\times10^{8}$ cm$^{-2}$ assuming $m^*=0.25m_0$. $m^*$ is an estimate since a complicated dispersion associated with the light-heavy hole bands mixing and the spin-orbit coupling~\cite{Winkler2003Spin} is not yet well-understood. On the other hand, the onset of an insulator as a function of $r_s$ can be obtained via the metal-to-insulator transition (MIT)~\cite{Kravchenko1995Scaling}. As shown in Fig.~\ref{fig:HIGFET}(c), $T$ dependence of $\rho(T)$, for $p$ from $7\times10^{8}$ to $5.6\times10^{9}$ cm$^{-2}$, yields a sign change in $d\rho/dT$ around a critical density of $p_c\sim4\times10^{9}$ cm$^{-2}$. Remarkably, $p_c$ corresponds to $r_s\sim 38$, the expected onset of a WC. Based on the RIP study, the 40-500 mK $T$ range covers from an intermediate phase to a liquid phase. $\rho(T)$ follows a nonactivated power law down to $p=7\times10^{8}$\,cm$^{-2}$, as indicated via linear trends in the double-logarithmic scales, contrasting the hopping conduction~\cite{Mott1969Conduction,Efros1975Coulomb} found in more disordered systems. For details of the nonactivated behavior and its conversion to activated transport in response to an increase of disorder, refer to references~\cite{Huang2011Light,Huang2012Nonactivated}.

\begin{figure}[t]
\vspace{-0pt}
\includegraphics[totalheight=2.4in]{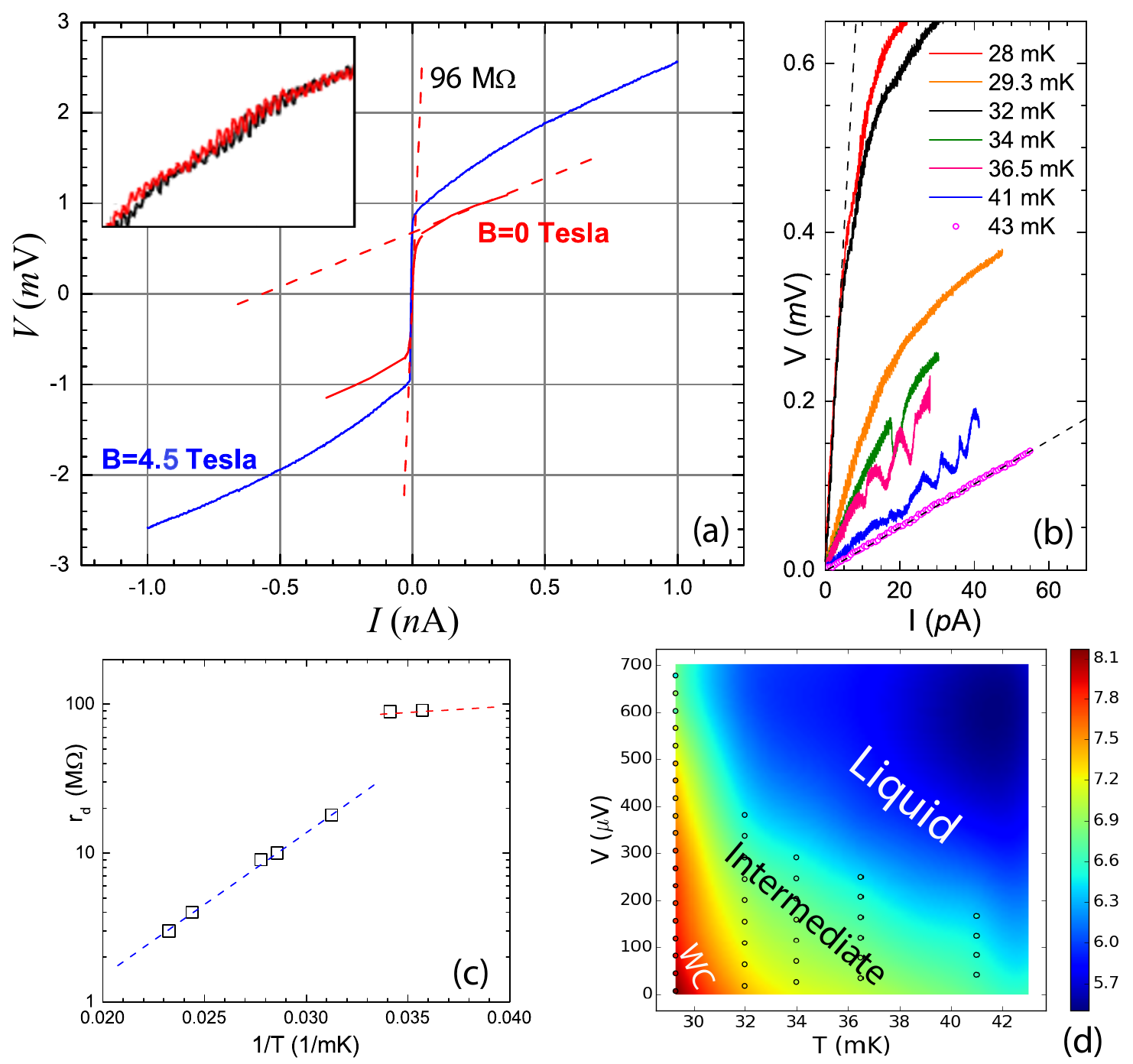}
\vspace{-10pt}
\caption{\label{fig:dc} (Color online) (a) DC-IV for $T=$28~mK. (b) IVs obtained at different $T$. (c) Discontinous $T$ dependence of maximum $r_d(T)$ on semi-logarithmic scales. (d) Colored contour phase diagram based on $\log_{10}{R_d}(T)$ values. Dashed lines are guide to the eye.}
\vspace{-15pt}
\end{figure}
Alternative to the DC voltage bias used for the RIP case, a current bias, with Keithley 6430 $f$A source, is employed with a sub-$\mu$V-resolution voltage sensing at an input impedance of $10^{16}\Omega$. Forcing a constant $I$ provides a more stringent probe to a pinned WC because it allows local effects such as melting to occur.  Nonlinear IV is expected to be weakened in an inferior pinning, i.e. in an intermediate/mixed phase (or a glass). On the contrary, a pinned rigorous WC should still exhibit a sharp threshold just as for the voltage-driven scenario, even though the voltage reading above the threshold is an overestimated cordal resistance (described in Ref.~\cite{butcher2013physics}) due to the separation of current and voltage channels. 

The following DC results are for $p=2.8\times10^{9}$ cm$^{-2}$ (or $r_s\sim45$) measured between 28 to 45 mK. The (Fermi) wavelength $\lambda$ is $\approx400n$m, twice the carrier spacing. 
A pinned WC is confirmed by a sharp threshold IV obtained at 28 mK shown in Fig.\ref{fig:dc}(a). This first demonstration of rigorous pinning in a zero-$B$ is marked by a 90~M$\Omega/\square$ sub-threshold $r_d$ below a tiny $I_c\sim 4 p$A. The supra-threshold $r_d$ collapses one hundred times, slightly weaker than the RIP case, which is consistent with an overestimated $V$ mentioned above.   
$I_c$ corresponds to a smaller threshold field $E_c\sim 4m$V/cm (or $\sim10^{-10}$ V$/a_B$), yielding a slightly larger single particle potential energy of $\sim eE_c a\sim0.04\mu eV\sim0.46$ mK due to the larger $a\sim 100$~nm. Setting $NeE_c=\kappa a$ as shown earlier, one obtains $N\sim5\times10^5$, corresponding to a WC on a substantial scale of $\xi\sim100\mu$m. The dominating potential energy $U\sim20me$V $>E_{ee}$ is consistent with a crystal. For a consistency check, the same setup is used to measure the RIP and the result is shown as the blue curve. The power dissipation is $\leq2\times10^{-16}$ Watts, ruling out appreciable Joule heating. 

Pressure ($E$)-driven melting occurs in a similar fashion to RIP except for slight oscillations of $r_d$ in the intermediate phase [inset of Fig.~\ref{fig:dc}(a)]. The IV curves for increasing $T$ shown in Fig.~\ref{fig:dc}(b) display an involvement from rigorous to soft pinning and then to an absence of pinning, consistent with a two-stage melting. Liquefaction occurs at 42~mK, noticeably lower than the RIP. This is likely due to stronger effects of quantum fluctuations at small wavevectors and disorder fluctuations due to lack of screening in such a dilute limit. Oscillations in $r_d$ are stronger in the $T$-driven intermediate phase than in the $E$-driven intermediate phase. It is possibly related to the formation of stripes with long-range orientational order~\cite{Nelson1979Dislocation} that vary with increasing drive, as seen in electrons on a helium surface~\cite{WC_helium}. Another possibility is that small $T_l$ facilitates a melting and re-crystallization of pinned WC domains, instead of or in addition to shearing, when driven across the fixed points of disorder~\cite{Normand1992Pinning}. 

Melting transition probed by maximum $r_d$ (at minimal bias) is shown in Fig.~\ref{fig:dc}(c) where a piecewise behavior, slight $T$ dependence below $T_m$ and the exponential $T$ dependence above $T_m$, is identical to the RIP results. Across $T_m\sim 30$~mK, the energy (or $\xi$) discontinuity occurs more abruptly than the RIP, in a manner agreeing with a recent quantum Monte Carlo simulation for a first order WC-intermediate phase transition mediated by a discontinuous jump in topological defects~\cite{qmHexatic}. The first-order nature could be a unique feature of the melting of a quantum system. A phase diagram is shown in Fig.~\ref{fig:dc}(d).



To summarize, a complete melting of a quantum WC has been captured as a two-stage SLT similar to the KT model. Distinctions between a WC and an intermediate phase are evidenced by the disappearance of $E_c$, a drastically different $T$ dependence corresponding to a discontinuity in the internal energy (or $\xi$), and an enormous $\xi$ for a pinned WC. They strongly support a first-order WC-intermediate phase transition which is qualitatively agreeable with Ref.~\cite{hexatic_ceperley}, instead of the second order KT model via the unbinding of dislocations. Effective cooling and low disorder are crucial to the formation of a WC, which is proven by rigorously pinned modes that survived stringent DC tests. $T_m$ is well below $T_{cm}$, indicating strong effects from quantum fluctuations and system disorders that require further understanding. 

We acknowledge the support of this work from NSF under DMR-1410302. The work at Princeton was partially funded by the Gordon and Betty Moore Foundation through Grant GBMF2719, and by the National Science Foundation MRSEC-DMR-0819860 at the Princeton Center for Complex Materials.

\bibliography{refs.bib}

\end{document}